Six postscript figures are appended at the end.
Each is preceeded by a line like ---- Fig. x cut here -----
and ends with "showpage".
\documentstyle[preprint,revtex,eqsecnum]{aps}

\begin{document}
\draft
\begin{title}
\begin{center}
Effect of three-particle correlations\\
in low dimensional Hubbard models
\end{center}
\end{title}
\author{Theodore C. Hsu${}^{(1),(2)}$\cite{byline}
and Beno\^it Dou\c{c}ot${}^{(2)}$\cite{dbyline}}
\begin{instit}
${}^{(1)}$AECL Research, Chalk River Laboratories, Chalk River,
Ontario,
Canada K0J 1J0
\end{instit}
\begin{instit}
${}^{(2)}$Centre des Recherches sur
les Tr\`es Basses Temp\'eratures,\\
BP 166, 38042 Grenoble, France
\end{instit}
\receipt{}
\begin{abstract}
A simple approximation which captures some
non-perturbative aspects of the one electron Green function
of strongly interacting Fermion systems is developed. It
provides a way to go one step beyond the usual dilute limit
since particle-particle as well as particle-hole scattering
are treated on the same footing. Intermediate
states are constrained to contain only one particle-hole excitation
besides the incoming particle. The Faddeev equations
resulting from an exact treatment of this three-body problem
are investigated. In one dimension the method is able
to show spin and charge decoupling, but does not reproduce
the exact nature of power-law singularities.
\end{abstract}
\narrowtext

\section{Introduction}

Since the discovery of high temperature superconductors,
much work from both experimental and theoretical sides has
suggested that Fermi-liquid theory does not provide a good
description of the remarkable normal state properties of
these materials. In particular, Anderson \cite{ANDERSONA}
has proposed that a better framework is to be found in some
two dimensional generalization of the Luttinger liquid,
which has been identified by Haldane \cite{HALDANE}
as a low energy effective
theory for many interacting-Fermion systems in one dimension.
For example, recent measurements of
the temperature dependence of the
Hall angle in high $T_{c}$ cuprates have been interpreted along
this line \cite{HALL}. However,
it has not yet been possible to derive
such a picture from a first principles calculation.
A suggestive investigation of
two-particle scattering in two dimensions in the presence
of the Fermi sea has lead Anderson to claim that a
finite phase shift and a singular interaction
function $f_{k,k^{\prime}}$ exist even in the dilute
limit \cite{ANDERSONB}. But if the same particle-particle ladder
is used in a more conventional many-body calculation,
it has been shown by Engelbrecht and Randeria \cite{RANDERIA}
and Fukuyama {\it et al.} \cite{FUKUYAMA}
that the imaginary part of the self-energy is not
singular enough in this T-matrix approximation to avoid
Fermi-liquid behavior.
In fact, the two approaches
may not contradict each other. Anderson suggests that an
extra particle added to the system induces a shift of
all the momenta of the particles already present in a similar
way as the deep level in the X-ray edge problem. The present
paper is an attempt to connect this physical picture to some
microscopic perturbative calculations for the repulsive Hubbard
model.

In our approach, we have been guided by
the idea already stressed in the
work by Abrikosov on the Kondo problem \cite{ABRIKOSOV},
and Nozi\`eres {\it et al.}
on the X-ray edge problem \cite{NOZIERES1},
namely that both particle and hole
scattering channels with the local impurity
lead to similar singularities
which may cancel partially. It is then
crucial to treat both in an unbiased
way. This is for instance a key feature
of the parquet summation, which was
developed in references \cite{ABRIKOSOV} and \cite{NOZIERES1}.
However, this method is
not very easy to implement for many body systems,
where the angular dependence
of interaction vertices is in general not
described by a small set of relevant
parameters (except in one dimension). A simplification
occurs if we work in
a truncated Hilbert space with only one particle-hole excitation
on top of
a local potential (X-ray edge problem), a spin (Kondo problem), or an
added electron (Hubbard model). This approach is variational in spirit
and amounts to treating, exactly,
three-particle correlations amongst the
particle-hole pair and the analogue of a
local scatterer. It has been used
in the context of Kondo problems to study various properties
\cite{YOSIDA,VARMA1}, and for the Hubbard
model in the special case of a
single overturned spin \cite{IGARASHI}.
It has even been suggested that
three-particle correlations between a spin flip
and two holes may provide
a way to reduce the ground state magnetization away
from its maximal value
in the infinite-U Hubbard model at any
density \cite{ANDREI}. Three-body
correlations are also investigated as a possible
microscopic mechanism
leading to a marginal Fermi-liquid \cite{VARMA2}.

In this paper then, we shall concentrate on
the effect of
three-particle correlations on the self-energy of
an electron. We consider a variational state which consists
of one electron plus at most one extra particle-hole pair
excitation on top of an otherwise rigid Fermi sea.
The scattering in the particle-particle
and particle-hole channels are treated on an equal footing.
In this sense we attempt to go beyond references \cite{RANDERIA}
and \cite{FUKUYAMA} who considered only two-particle scattering.
We shall be interested in seeing whether the Fermi-liquid
starting point is valid or not.
Exact solutions are available for the Luttinger model and
the X-ray edge problem and we have decided to
compare the electron's Green function in the three-body
approximation to that of the exact solution for these systems.
In section 2 we shall discuss the formalism leading to
the Faddeev integral equations \cite{FADDEEV}.
We have re-formulated these
in order to yield directly the self-energy. Sections 3 and 4
will describe results for the one and two branch Luttinger
models respectively and section 5 will describe results
for the X-ray problem.

\section{Faddeev Equations}
We begin with the definition of the electron Green function,
\begin{equation}
G({\bf q},\omega) = \langle N\mid
c_{{\bf q}\uparrow} [\omega - H +i\delta]^{-1}
c^{\dagger}_{{\bf q}\uparrow} +
c^{\dagger}_{{\bf q}\uparrow} [\omega + H -i\delta]^{-1}
c_{{\bf q}\uparrow}\mid N\rangle
\end{equation}
where the Hamiltonian is
\begin{equation}
H = \sum_{{\bf k}} \epsilon_{{\bf k}}
c^{\dagger}_{{\bf k}\sigma}
c_{{\bf k}\sigma}
+ {U\over N_{s}}\sum_{{\bf k},{\bf k}^{\prime},{\bf p}}
c^{\dagger}_{{\bf k}+{\bf p}\uparrow}
c_{{\bf k}\uparrow}
c^{\dagger}_{{\bf k}^{\prime}-{\bf p}\downarrow}
c_{{\bf k}^{\prime}\downarrow}
- E_{0}
\quad .
\label{HAMILTONIAN}
\end{equation}
We shall consider a variational state where
$\mid N\rangle$ is the non-interacting
Fermi sea and the Hamiltonian
is allowed to create at most one excited particle-hole pair.
The corresponding diagrams contributing to the Green function
are shown in Fig. \ref{DIAGRAM}.
We take $q > k_{F}$ and
$E_{0}$ is chosen so that $\langle N\mid H\mid N\rangle = 0$.
In this case, defining
\begin{equation}
\phi\equiv
\langle N\mid c_{{\bf q}\uparrow} [\omega - H +i\delta]^{-1}
c^{\dagger}_{{\bf q}\uparrow}\mid N\rangle
\end{equation}
and
\begin{equation}
\Phi({\bf k},{\bf k}^{\prime})\equiv
\langle N\mid c_{{\bf q}\uparrow} [\omega - H +i\delta]^{-1}
c^{\dagger}_{{\bf k}\uparrow}
c^{\dagger}_{{\bf k}^{\prime}\downarrow}
c_{{\bf k}+{\bf k}^{\prime}-{\bf q}\downarrow}\mid N\rangle
\quad ,
\end{equation}
the equations of motion of these Green functions
truncating terms involving more than three particles are
\begin{equation}
[\tilde{\omega} - \epsilon_{{\bf q}}]\phi
- {U\over N_{s}}\sum_{{\bf k},{\bf k}^{\prime}}
\Phi({\bf k},{\bf k}^{\prime})
= 1
\label{MOTIONA}
\end{equation}
and
\begin{equation}
[\tilde{\omega} - \epsilon({\bf k},{\bf k}^{\prime})]
\Phi({\bf k},{\bf k}^{\prime}) - {U\over N_{s}}\phi
-{U\over N_{s}}\sum_{{\bf k}^{\prime\prime}}
\Phi({\bf k}^{\prime\prime},{\bf k}+
{\bf k}^{\prime}-{\bf k}^{\prime\prime})
+{U\over N_{s}}\sum_{{\bf k}^{\prime\prime}}
\Phi({\bf k}^{\prime\prime},{\bf k}^{\prime})
= 0
\label{MOTIONB}
\end{equation}
where $\tilde{\omega} = \omega - U(N_{\downarrow}/N_{s})$ and
$\epsilon({\bf k},{\bf k}^{\prime}) = \epsilon_{{\bf k}}
+ \epsilon_{{\bf k}^{\prime}}
- \epsilon_{{\bf k} + {\bf k}^{\prime} - {\bf q}}$.
The momentum restrictions in the sum are such that the
arguments of $\Phi ({\bf k}_{1},{\bf k}_{2})$ always satisfy
${\bf k}_{1}\neq {\bf q}$; $k_{1},k_{2} > k_{F}$
and $|{\bf k}_{1}+{\bf k}_{2}-{\bf q}| < k_{F}$.

We shall find it convenient to define the following functions:
\begin{equation}
J_{1}({\bf K}) = N_{s}^{-1}\sum_{{\bf k}^{\prime}}
\Phi({\bf k}^{\prime},{\bf K}-{\bf k}^{\prime})
\end{equation}
and
\begin{equation}
J_{2}({\bf K}) = N_{s}^{-1}\sum_{{\bf k}^{\prime}}
\Phi({\bf k}^{\prime},{\bf K})
\end{equation}
which correspond roughly to T-matrices for the particle-particle and
particle-hole scattering channels respectively.
{}From the equations of motion for $\phi$ we may see that
$J_{1}$ is related to the self-energy by
\begin{equation}
\Sigma({\bf q},{\tilde\omega}) = U^{2}\sum_{{\bf K}}J_{1}({\bf K})
\quad .
\end{equation}

{}From the equation of motion for $\Phi$ we see that $J_{1}$
and $J_{2}$ satisfy
the coupled integral equations
\begin{equation}
J_{1}({\bf K}) =
N_{s}^{-1}
{{F_{1}({\bf K})}\over{1 - UF_{1}({\bf K})}}
- {U\over N_{s}}{{1}\over{1 - UF_{1}({\bf K})}}
\sum_{{\bf k}^{\prime}}
{
	{
		J_{2}({\bf K}-{\bf k}^{\prime})
	}
\over
	{
		{\tilde\omega} - \epsilon({\bf k}^{\prime},
		{\bf K}-{\bf k}^{\prime})
	}
}
\end{equation}
and
\begin{equation}
J_{2}({\bf K}) =
N_{s}^{-1}
{{F_{2}({\bf K})}\over{1 + UF_{2}({\bf K})}}
+ {U\over N_{s}}{{1}\over{1 + UF_{1}({\bf K})}}
\sum_{{\bf k}^{\prime}}
{
	{
		J_{1}({\bf K}+{\bf k}^{\prime})
	}
\over
	{
		{\tilde\omega} - \epsilon({\bf k}^{\prime},{\bf K})
	}
}
\quad .
\end{equation}
In the summations the same momentum restrictions as the equations
of motion Eqs. (\ref{MOTIONA}) and (\ref{MOTIONB})
are in effect. $F_{1}$ and $F_{2}$ are defined by
\begin{equation}
F_{1}({\bf K}) = N_{s}^{-1}\sum_{{\bf k}^{\prime}}
{
	{1}
\over
	{{\tilde\omega}-\epsilon({\bf k}^{\prime},
	{\bf K}-{\bf k}^{\prime})}
}
\end{equation}
and
\begin{equation}
F_{2}({\bf K}) = N_{s}^{-1}\sum_{{\bf k}^{\prime}}
{
	{1}
\over
	{{\tilde\omega}-\epsilon({\bf k}^{\prime},{\bf K})}
}
\quad .
\end{equation}

The integral equations can be combined into a single one for
$J_{1}$ which is the final equation to be solved.
\begin{eqnarray}
&J_{1}({\bf K}) =
{1\over N_{s}^{2}}{{1}\over{1 - UF_{1}({\bf K})}}
\sum_{{\bf k}^{\prime}}
{
	{
		1
	}
\over
	{
		({\tilde\omega} - \epsilon({\bf k}^{\prime},
		{\bf K}-{\bf k}^{\prime}))
		(1 + UF_{2}({\bf K}-{\bf k}^{\prime}))
	}
}
\nonumber
\\
&- {U^{2}\over N_{s}^{2}}{{1}\over{1 - UF_{1}({\bf K})}}
\sum_{{\bf k}^{\prime\prime}}\sum_{{\bf k}^{\prime}}
{
	{
		J_{1}({\bf k}^{\prime})
	}
\over
	{
		({\tilde\omega} -
		\epsilon({\bf k}^{\prime\prime},
		{\bf K}-{\bf k}^{\prime\prime}))
		(1 + UF_{2}({\bf K}-{\bf k}^{\prime\prime}))
		({\tilde\omega} -
		\epsilon({\bf k}^{\prime}-{\bf K}
		+{\bf k}^{\prime\prime},
		{\bf K}-{\bf k}^{\prime\prime}))
	}
}
\quad .
\end{eqnarray}

The technique of truncating the Green function equations of motion
at the four point level and finding closed equations for them
is not limited to well defined particles and is not new.
It has been applied \cite{IRKHIN} to
Green functions for the Hubbard `X' operators \cite{HUBBARD}.

\section{One--branch Luttinger model}
In this section we present the exact analytical
solution of the integral equation for a one branch
Luttinger model (i.e. with right movers only). The total momentum of
the three particles is $q>k_{F}$ and the energy dispersion is defined
by $\epsilon_{k} = v_{F}(k-k_{F})$. The simplification which allows
an analytical solution is the fact that $\epsilon(k,k^{\prime})
\equiv \epsilon_{k} + \epsilon_{k^{\prime}}
- \epsilon_{k+k^{\prime}-q}
= v_{F}(q-k_{F})$.
Thus we have
\begin{equation}
F_{1}(K) =
{
	{K-2k_{F}}
\over
	{{\tilde\omega}-v_{F}(q-k_{F})}
}
\end{equation}
and
\begin{equation}
F_{2}(K) =
{
	{q-K}
\over
	{{\tilde\omega}-v_{F}(q-k_{F})}
}
\quad .
\end{equation}

Defining ${\tilde q}\equiv q-k_{F}$,
\begin{eqnarray}
j_{1}(k) \equiv& ({\tilde\omega} - v_{F}{\tilde q})^{-1}J_{1}(k)\\
f_{1}(k) \equiv& ({\tilde\omega} - v_{F}{\tilde q})^{-1}
	[1 - UF_{1}(k)]^{-1}\\
f_{2}(k-k^{\prime\prime}) \equiv& ({\tilde\omega} - v_{F}{\tilde q})^{-1}
	[1 + UF_{2}(k-k^{\prime\prime})]^{-1}
\end{eqnarray}
we absorb a factor of $N_{s}$ into $J_{1}$ and go to
the continuum  obtaining
\begin{equation}
j_{1}(k) = f_{1}(k)\int_{k_{F}}^{k-k_{F}}dk^{\prime\prime}
f_{2}(k-k^{\prime})
\left[
	1 - U^{2}\int_{k-k^{\prime\prime}+k_{F}}^{q+k_{F}}
	dk^{\prime}j_{1}(k^{\prime})
\right]
\quad .
\end{equation}

There is, in fact, no small parameter in this integral
equation, even when U is small.
Thus it cannot be solved perturbatively.
This can be demonstrated explicitly by defining
$j(k)\equiv {\tilde q}U^{2}j_{1}(k)$ and
$x\equiv ({\tilde\omega} - v_{F}{\tilde q})/U{\tilde q}$.
Rescaling all momenta with respect to ${\tilde q}$
and setting $k_{F}=0$ we have
\begin{equation}
j(k) = {1\over{x-k}}\int_{0}^{k}
dk^{\prime\prime}
{{1}\over{x+1-k+k^{\prime\prime}}}
\left[
	1 - \int_{k-k^{\prime\prime}}^{1}
	dk^{\prime}
	j(k^{\prime})
\right]
\quad .
\end{equation}

Changing the order of integration allows us
to perform one of the nested integrals.
Using
\begin{equation}
\int_{0}^{k}dk^{\prime\prime}
\int_{k-k^{\prime\prime}}^{1}dk^{\prime}
=
\int_{0}^{k}dk^{\prime}
\int_{k-k^{\prime\prime}}^{k}dk^{\prime\prime}
+
\int_{k}^{1}dk^{\prime}
\int_{0}^{k}dk^{\prime\prime}
\end{equation}
we find
\begin{eqnarray}
&j(k) = {1\over{x-k}}\times \nonumber \\
&\left[
	\ln
	\left(
		{{x+1}\over{x+1-k}}
	\right)
	-
	\int_{0}^{k}dk^{\prime}j(k^{\prime})
	\ln
	\left(
		{{x+1}\over{x+1-k^{\prime}}}
	\right)
	-
	\int_{k}^{1}dk^{\prime}j(k^{\prime})
	\ln
	\left(
		{{x+1}\over{x+1-k}}
	\right)
\right]
\quad .
\end{eqnarray}

The solution proceeds by conversion into
a differential equation.
Multiplying by $x-k$ and differentiating with respect
to $k$ we have
\begin{equation}
{d\over{dk}}
\left[
(x-k)j(k)
\right]
=
{{1}\over{x+1-k}}
-
{{1}\over{x+1-k}}
\int_{k}^{1}dk^{\prime} j(k^{\prime})
\quad .
\end{equation}
Multiplying by $x+1-k$ and differentiating,
\begin{equation}
{d\over{dk}}
\left[
(x+1-k)
((x-k)j^{\prime}(k)
-
j(k))
\right]
=
j(k)
\quad .
\label{DIFF}
\end{equation}
This can be integrated because the integral equation
gives us the two boundary conditions
\begin{equation}
j(0)=0
\end{equation}
and
\begin{equation}
x(x+1)j^{\prime}(0)
=
1 - \int_{0}^{1}dk j(k)
\equiv
1 - (\Sigma(x)/xU{\tilde q})
\quad .
\end{equation}
Integration of Eq. (\ref{DIFF}) results in an equation for
$j(k)$ in terms of $x$,$k$, and $j^{\prime}(0)$.
Integrating over $k$ from $k=0$ to $k=1$ eliminates
$j(k)$ and $k$ and results in
an equation for the self-energy in terms of $x$
whose solution is (adding in the imaginary parts now)
\begin{equation}
\Sigma(x) =
xU{\tilde q}{\cal A}/(1 + {\cal A})
\end{equation}
where
\begin{equation}
{\cal A} =
x^{2}
\ln
\left|
{{x^{2}}
\over
{1-x^{2}}}
\right|
-1
-i\pi x|x|\theta(x+1)\theta(1-x)
\quad .
\end{equation}
Substituting back into the electron Green function
gives the very simple result
\begin{equation}
G(x) = x
\ln
\left|
{{x^{2}}
\over
{1-x^{2}}}
\right|
-
i\pi |x|\theta(x+1)\theta(1-x)
\label{SOLUTIONA}
\end{equation}
where $\theta(x)$ is the usual step function.
Re-inserting the original units,
\begin{equation}
{\rm Im} G({\tilde\omega},{\tilde q}) =
\pi
\left|
{
{{\tilde\omega}-v_{F}{\tilde q}}
\over
{U{\tilde q}}
}
\right|
,\quad
v_{F}{\tilde q} - U{\tilde q}
< {\tilde\omega} <
v_{F}{\tilde q} + U{\tilde q}
\label{SOLUTIONB}
\end{equation}
The exact Green function is \cite{VOIT}
\begin{equation}
{\rm Im} G({\tilde\omega},{\tilde q}) =
\left[
(U{\tilde q})^{2} - ({\tilde\omega}-v_{F}{\tilde q})^{2}
\right]^{-1/2}
,\quad
v_{F}{\tilde q} - U{\tilde q}
< {\tilde\omega} <
v_{F}{\tilde q} + U{\tilde q}
\label{ONEB_EXACT}
\end{equation}

Our result is compared to the exact result in Fig. \ref{ONEBRANCH}.
The spectral weight
from the three-body approximation is non-zero over the same
energy range as the exact solution and has roughly the
same shape.
Although the two have maxima in the same place,
the three-body approximation is not able to reproduce
the square root singularities of the exact solution.
Nevertheless spin-charge separation is observed.
The
charge velocity $v_{c} = v_{F} + U$ and the spin velocity
$v_{s} = v_{F} - U$ agree with the exact result.
{}From the integral equations one can see that
the `charge' peak comes from $[1 - UF_{1}]^{-1}$ and
the `spin' peak comes from $[1 + UF_{2}]^{-1}$.

It should be remarked that the one branch model has the
same ground state as non-interacting Fermions.
Thus spin-charge separation can co-exist with
a Fermi-liquid.
The range over which
Im~G is non-zero is proportional to $q-k_{F}$
and this goes to zero as $q\rightarrow k_{F}$.
Thus the three-body approximation is self-consistent
with the assumption of a rigid Fermi
surface background. This can be understood by
noting that it is impossible
to create particle-hole excitations from the ground state
and conserve momentum. It is this feature of the one-branch
model which renders the three-body approximation
tractable. Below, in the two branch model, this phenomenon
does not occur.

Finally we would like to emphasize the non-perturbative
nature of the integral equation. This feature may help in
shedding light on whether or not the two-dimensional
Hubbard model has a Fermi-liquid ground state.
Initial calculations indicate that $[1 - UF_{1}]^{-1}$ and
$[1 + UF_{2}]^{-1}$ are less singular in two dimensions
and almost certainly will not lead to singular
behaviour on their own. We hope to discuss this in detail
in a future publication.

\section{Two branch Luttinger model}
We consider a spinless Luttinger model for which only
the interaction between opposite branches is retained.
Again we begin with a particle on the right
moving branch but allow the creation of a single particle-hole
pair on the left moving branch.
In this case the integral equation is
(absorbing
a factor of $N_{s}$ into $J_{1}$),
\begin{eqnarray}
J_{1}(K) =
{{1}\over{1 - UF_{1}(K)}}
\sum_{k^{\prime\prime} = K+k_{F}}^{D}
{
	{
		1
	}
\over
	{
		\left(
		{\tilde\omega}
		- \epsilon(k^{\prime\prime},K-k^{\prime\prime})
		\right)
		\left(
		1 + UF_{2}(K-k^{\prime\prime})
		\right)
	}
}
\nonumber
\\
- {{U^{2}}\over{1 - UF_{1}(K)}}
\left[
\sum_{k^{\prime} = q-k_{F}}^{K}
\sum_{k^{\prime\prime} = K+k_{F}}^{D}
+
\sum_{k^{\prime} = K}^{D-k_{F}}
\sum_{k^{\prime\prime} = K+k_{F}}^{D+K-k^{\prime}}
\right]
\nonumber \\
\times
{
	{
		J_{1}(k^{\prime})
	}
\over
	{
		({\tilde\omega} -
		\epsilon(k^{\prime\prime},K-k^{\prime\prime}))
		(1 + UF_{2}(K-k^{\prime\prime}))
		({\tilde\omega} -
		\epsilon(k^{\prime}-K+k^{\prime\prime},
		K-k^{\prime\prime}))
	}
}
\end{eqnarray}
where the kinetic energy
$\epsilon(k^{\prime},k-k^{\prime})
 = (2k^{\prime} - q - k_{F})v_{F}$ (It doesn't depend on $k$
because the total momentum must sum to $q$.),
\begin{equation}
F_{1}(k) = \sum_{k^{\prime} = k+k_{F}}^{D} [{\tilde\omega}-
\epsilon(k^{\prime},k-k^{\prime})]^{-1}
\end{equation}
and
\begin{equation}
F_{2}(k) = \int_{k^{\prime} = -k+{\tilde q}}^{D} [{\tilde\omega}-
\epsilon(k^{\prime},k)]^{-1}\quad .
\end{equation}
D is a momentum cutoff.

We were not able to solve this analytically,
mostly because of the intractability of
integrals of the form $\int dz{\rm ln}(z)/(z+a)$,
and therefore attempted a numerical solution by discretization.
In order to control the divergences of the free particle
poles we introduced artificial widths. That is, we replaced
\begin{equation}
{1\over{{\tilde\omega}-{\tilde q}}}\rightarrow
{1\over{{\tilde\omega}-{\tilde q}+i\delta}}
\end{equation}
We checked that our results did not depend
significantly on the value of $\delta$.
The method of numerical solution was checked against
the analytical solution, Eq. (\ref{SOLUTIONA})
for the one-branch model, to confirm
its accuracy.

The exact Green function for this problem is \cite{VOIT}
\begin{eqnarray}
&{\rm Im} G({\tilde q},{\tilde w}) =
-
{
\pi
\over
{2D^{\prime}v\Gamma(a)^{2}}
}
\nonumber\\
&\times
\left[
\theta({\tilde\omega}-v{\tilde q})
\gamma(a,y_{+})
e^{-y_{-}}
y_{-}^{a-1}
+
\theta({-\tilde\omega}-v{\tilde q})
\gamma(a,-y_{+})
e^{y_{-}}
(-y)_{-}^{a-1}
\right]
\end{eqnarray}
where $y_{\pm} = ({\tilde\omega}\pm v{\tilde q})/2D^{\prime}v$,
$v = \sqrt{v_{F}^{2}-U^{2}}$
and
\begin{equation}
a = {1\over 4}(\sqrt{{v_{F}-U}\over{v_{F}+U}}
+
\sqrt{{v_{F}+U}\over{v_{F}-U}}
-
2)
\quad .
\end{equation}
The parameter $D^{\prime}$ is a momentum cutoff. In a perturbative
expansion of the integral equations of the three-body approximation
the parameter $({\tilde\omega} - v_{F}{\tilde q})/2Dv_{F}$
appears.
In order to compare the three-body
results to the exact results we need definite values and it seems
reasonable to take $D^{\prime}=D$ for this purpose
so that the three-body solution and the exact solution
have the same dimensionless parameter
$({\tilde\omega} - v_{F}{\tilde q})/2Dv_{F}$.
${\rm Im} G({\tilde q},{\tilde\omega})$
(for the exact and three-body solution)
as a function of energy with fixed
total momentum are superimposed in Fig. \ref{TWOBRANCH}.
Energies are normalized by ${\tilde q}v_{F}$.
We chose a value $U/v_{F} = 0.3$ which is large enough so that
the effects of $U$ are apparent but not so large that
the integral equations do not converge. The
cutoff is $(D-k_{F})/{\tilde q} = 3$ and is not a
sensitive parameter.

Some differences are immediately evident.
Firstly the exact solution has spectral weight at
${\tilde\omega} < -v{\tilde q}$
(which is not shown in the figure).
The three-body approximation misses this.
In order to describe this part of the spectrum
we would have to include
pre-existing particle and hole
pair excitations in the N-particle ground state
whereas
the three-body approximation assumes that the background is a rigid
Fermi sea. A background of pre-excited particle-hole pairs
would require consideration of a minimum of five bodies.
The second difference is that the spectral weight of
the exact solution is extremely concentrated
just above the maximum of the spectral function. That is
because, for small U, Im~G diverges with a
negative exponent only slightly greater
than $-1$. The exact solution displays an interesting characteristic.
For energies $|{\tilde\omega}| < v{\tilde q}$ the spectral weight is
exactly zero. We do not know how to explain this in terms of
the underlying electrons.
This feature is not present in the three-body approximation.
Nevertheless the spectral function in that case is asymmetric.
In Fig. \ref{TWOBRANCH} the reflection of the spectral function
about its maximum is plotted to bring out the asymmetry.
We see that the asymmetry is in the right direction. That is,
the spectral weight is higher for energies above the maximum.
By looking at the numerical solution integral equations
it is possible to tell that a pole in $[1 + F_{2}]^{-1}$
(the particle - hole scattering channel)
is the cause of this asymmetry.

Another feature which is reproduced qualitatively is the
negative energy shift of the peak position (real part of the
self-energy). In the exact solution this manifests itself
in a downward renormalization of the Fermi velocity
$v = \sqrt{v_{F}^{2} - U^{2}}$. In Fig. \ref{SHIFT}
the shifts in the maxima of Im G as a function
of U are plotted. It is interesting to note that the
three-body approximation follows the square root behaviour
of the exact solution quite accurately except for a factor of
two.

To summarize, the three-body approach reproduces
qualitatively the real part of the self-energy
and the asymmetry of the spectral function.
Although it is not shown in Fig. \ref{TWOBRANCH},
in the three-body approximation the spectral weight
of an electron with momentum $q>k_{F}$ stretches into
negative energy, with a width proportional to $q-k_{F}$.
However, the integrated spectral weight for negative
energies seems to be independent of $q-k_{F}$ at
small $q-k_{F}$ and amounts to a few percent of the
total spectral weight (for the given choice of parameters):
comparable to that of the exact solution.
This flow of spectral weight can be interpreted by
saying that the three-body approximation is
inducing some correlations which tend to improve
the trial ground state.
This confirms the idea that correcting the N-particle
ground state by coherent particle-hole
pair excitations
would yield a Green function more like
the exact solution.

\section{X-ray problem}

As discussed in the introduction, our goal was to set up
an approximation scheme which treats particle-particle
and particle-hole interactions on the same footing.
Formally, this idea was first
discussed in condensed matter physics
for the Kondo problem \cite{ABRIKOSOV}, and for the
simpler X-ray edge problem \cite{NOZIERES1}. In this section, we present
the results of our simple three-body type of
approach for the X-ray edge problem.
At this point, it would be helpful to characterize better the
expected limitations of
our method.
The quantity of
interest here is the overlap between the wave function
of the conduction electron system,
after the sudden turning on of a localized
potential, and the unperturbed Fermi sea.
Going beyond the first calculation
by Mahan \cite{MAHAN}, Nozi\`eres and coworkers have shown that
the deep level
Green's function defined as above decays
as a power-law function of
time \cite{NOZIERES2}, with an exponent proportional
to the square of the
phase shift at the Fermi energy due to the local potential.
This result has
been interpreted in terms of Anderson's
orthogonality catastrophe \cite{CATASTROPHE,HOPFIELD}, and
rederived in the simpler language
of Tomonaga bosons \cite{SCHOTTE}. As stressed by many
authors, the infrared singularity arises from
the excitation of a logarithmically divergent number of
Tomonaga bosons. However, the average
number of such emitted bosons increases only logarithmically
with time, which
might enable us to restrict ourselves to the subspace
with no more than one excited boson, at least for times shorter than
$\frac{1}{W} \exp{(W/2V^{2})}$, where W is the conduction electron
bandwidth and V is the strength of the localized potential.

The Hamiltonian for the X-ray problem is
\begin{equation}
H = {\cal E}_{0}bb^{\dagger}
+ \sum_{\bf k}\epsilon_{\bf k}
a^{\dagger}_{\bf k}a_{\bf k}
- E_{0}
+ bb^{\dagger}\sum_{{\bf k}{\bf k}^{\prime}}
V_{{\bf k}^{\prime},{\bf k}}
a^{\dagger}_{{\bf k}^{\prime}}a_{\bf k}\quad .
\end{equation}
$a_{k}$ is a spinless Fermion annihiliation operator and
${\cal E}_{0}$ is the energy of the deep hole.
As in the Hamiltonian \ref{HAMILTONIAN}, $E_{0}$ is chosen so that the
unperturbed Fermi sea with the deep level occupied
is defined to have zero energy.
Let $|N\rangle$ be the unperturbed Fermi sea and
$|0\rangle \equiv b|N\rangle$ be the unperturbed
sea with the deep hole.
We shall be interested in the Green function
\begin{equation}
{\cal G}(t>0) = -i\langle N|e^{iHt}b^{\dagger}e^{-iHt}b|N\rangle
= -ie^{-i{\cal E}_{0}t}\langle 0|e^{-iHt}|0\rangle
\end{equation}

We again set up the three-body
approximation using the equation of motion formalism.
Defining $|t\rangle \equiv \exp{(-iHt)}|0\rangle$,
we make the approximation
\begin{equation}
|t\rangle \cong \phi(t)|0\rangle +
\sum_{k>k_{F}}\sum_{k^{\prime}<k_{F}}
\Phi({\bf k},{\bf k}^{\prime},t)
a^{\dagger}_{\bf k}a_{{\bf k}^{\prime}}
|0\rangle
\end{equation}
with the truncated equations of motion
\begin{equation}
i{{\partial\phi}\over{\partial t}}(t) =
{\cal E}_{0}\phi(t) +
\sum_{k,k^{\prime}} V_{k^{\prime},k}
\Phi(k,k^{\prime},t)
\end{equation}
and
\begin{eqnarray}
i{{\partial \Phi}\over{\partial t}}(k,k^{\prime},t) =
&({\cal E}_{0} + \epsilon_{k} - \epsilon_{k^{\prime}})
\Phi(k,k^{\prime},t)
+ \sum_{k^{\prime\prime}}V_{k,k^{\prime\prime}}
\Phi(k^{\prime\prime},k^{\prime},t)\nonumber\\
&- \sum_{k^{\prime\prime}}V_{k^{\prime\prime},k^{\prime}}
\Phi(k,k^{\prime\prime},t)
+ V_{k,k^{\prime}}\phi(t) \quad .
\end{eqnarray}
We proceed by expressing $\Phi({\bf k},{\bf k}^{\prime},t)$ in the basis
of single-particle eigenstates in the presence of scattering
$V_{k,k^{\prime}}$,
\begin{equation}
\Phi(k,k^{\prime},t) = \sum_{\alpha ,\beta}\Phi_{\alpha ,\beta}(t)
\phi^{p}_{\alpha}(k)
\phi^{h}_{\beta}(k^{\prime})
\end{equation}
in which case the second equation of motion may be written
\begin{equation}
i\sum_{\alpha ,\beta}
\phi^{p}_{\alpha}(k)
\phi^{h}_{\beta}(k^{\prime})
{\dot\Phi}_{\alpha ,\beta}(t)
=\sum_{\alpha ,\beta}
(E^{p}_{\alpha} + E^{h}_{\beta} + {\cal E}_{0})
\phi^{p}_{\alpha}(k)
\phi^{h}_{\beta}(k^{\prime})
\Phi_{\alpha ,\beta}(t)
+
V_{k,k^{\prime}}
\phi(t)\quad .
\end{equation}
Using the orthogonality of this basis we have
\begin{equation}
\Phi_{\alpha ,\beta}(t)
=
-i\int_{0}^{t}dt^{\prime}
\exp{
\left[
-i(E^{p}_{\alpha} + E^{h}_{\beta} + {\cal E}_{0})(t-t^{\prime})
\right]
}
\phi(t^{\prime})
\sum_{k,k^{\prime}}
\phi^{p*}_{\alpha}(k)
\phi^{h*}_{\beta}(k^{\prime})
V_{k,k^{\prime}}
\end{equation}

Combining this with the first equation of motion written in the
same basis and defining $\phi(t)\equiv\exp{(-i{\cal E}_{0}t)}
{\tilde\phi}(t)$,
\begin{equation}
{\tilde\phi}(t) = 1 + \int_{0}^{t}dt^{\prime}
\int_{0}^{t^{\prime}}dt^{\prime\prime}
\sum_{k,k^{\prime}}\sum_{q,q^{\prime}}
V_{k^{\prime},k}
V_{q,q^{\prime}}
G_{k,q}^{p}(t^{\prime}-t^{\prime\prime})
G_{k^{\prime},q^{\prime}}^{h}(t^{\prime}-t^{\prime\prime})
{\tilde\phi}(t^{\prime\prime})
\label{PHITILDE}
\end{equation}
where
\begin{equation}
G_{k,q}^{p,h}(t)
=
\sum_{\alpha}
e^{-iE_{\alpha}^{p,h}(t)}
\phi^{p,h}_{\alpha}(k)
\phi^{p,h*}_{\alpha}(q)
\end{equation}
are the particle and hole Green functions in the presence of
the deep hole.
Upon Fourier transforming Eq. (\ref{PHITILDE}) and solving
we find that
${\tilde\phi}(\omega) = i/\left(\omega - \Sigma(\omega)\right)$
where the self-energy $\Sigma$ is given by
\begin{equation}
\Sigma(\omega) = i\sum_{k,k^{\prime}}\sum_{q,q^{\prime}}
V_{k^{\prime},k}
V_{q,q^{\prime}}
\int {{d\omega^{\prime}}\over{2\pi}}
G_{k,q}^{p}(\omega^{\prime})
G_{k^{\prime},q^{\prime}}^{h}(\omega - \omega^{\prime})
\end{equation}
We take the continuum limit and treat the deep hole as
a point scatterer thus dropping momentum dependence in V.
The self-energy becomes
\begin{equation}
\Sigma = iV^{2}
\int {{d\omega^{\prime}}\over{2\pi}}
G^{p}(\omega^{\prime})
G^{h}(\omega - \omega^{\prime})
\end{equation}
where $G^{p,h}$ are now the on-site Green functions given
by
\begin{equation}
G^{p}(\omega) =
{
{
(1/2Dv_{F}){\rm ln}
\left[
(\omega - v_{F}k_{F} + i\delta)
/
(\omega - v_{F}D + i\delta)
\right]
}
\over
{
1 -
(V/2Dv_{F}){\rm ln}
\left[
(\omega - v_{F}k_{F} + i\delta)
/
(\omega - v_{F}D + i\delta)
\right]
}
}
\end{equation}
and
\begin{equation}
G^{h}(\omega) =
{
{
(1/2Dv_{F}){\rm ln}
\left[
(\omega + v_{F}k_{F} + i\delta)
/
(\omega - v_{F}D + i\delta)
\right]
}
\over
{
1 +
(V/2Dv_{F}){\rm ln}
\left[
(\omega + v_{F}k_{F} + i\delta)
/
(\omega - v_{F}D + i\delta)
\right]
}
}\quad .
\end{equation}
D is the usual momentum cutoff.

In Fig. \ref{IMG} we have plotted the imaginary part of the
particle and hole Green functions. Note the
weak logarithmic singularities which result from
single particle bound states. In Fig. \ref{SELF} is plotted
the resulting real and imaginary parts of the self-energy.
The critical features in the latter plot are the sharpness of the cusp
in ${\rm Im}\Sigma(\omega)$, and the steepness of the inflection
point of ${\rm Re}\Sigma(\omega)$ near $\omega\sim 0$.
As this numerical calculation shows, the weak logarithmic
divergences of the single particle spectral density do not
have a great effect on the self-energy.

In view of this, we proceed with an analytical calculation of the
quasiparticle residue which replaces the spectral functions
plotted in Fig. \ref{IMG} by the constant $-(2Dv_{F})^{-1}$
multiplied by appropriate step functions.
For small $\omega$,
\begin{eqnarray}
{\rm Re}\Sigma(\omega) \cong
(V/2Dv_{F})^{2}
&\left[
v_{F}(D-k_{F}){\rm ln}
\left(
{
{D-k_{F}}
\over
{2D}
}
\right)
+
v_{F}(D+k_{F}){\rm ln}
\left(
{
{D+k_{F}}
\over
{2D}
}
\right)
\right.\nonumber\\
&\left.
+
\omega
{\rm ln}
\left(
{
{|\omega|2Dv_{F}}
\over
{v_{F}(D-k_{F})v_{F}(D+k_{F})}
}
\right)
\right]
\quad .
\end{eqnarray}

The pole in the deep hole Green function
occurs at $\omega_{p}={\rm Re}\Sigma(\omega_{p})$
which is
\begin{equation}
\omega_{p} =
-
{
{V^{2}}
\over
{2Dv_{F}}
}
\left({\rm ln}2 + {\cal O}(k_{F}/D)^{2}\right)
\end{equation}
in the $k_{F} << D$ limit.
The residue of the deep level pole is
\begin{equation}
Z_{p} \equiv
\left[
1 - \left.{{\partial}\over{\partial\omega}}
{\rm Re}\Sigma(\omega)\right|_{\omega_{p}}
\right]^{-1}
\cong
1 - (V/2Dv_{F})^{2}
{\rm ln}
\left(
{
{v_{F}^{2}D^{2}}
\over
{V^{2}{\rm ln}2}
}
\right)
\quad .
\end{equation}
These expressions have been quantitatively checked with
a numerical calculation.

This calculation shows that the simple
three-body approach does not reproduce
the power-law singularity of the deep level
Green's function at low energy.
Instead of a power-law decay at long times, we obtain a long-lived
level, with a residue close to unity at small coupling. This result is
by itself not too surprising. It simply confirms the intuitive idea that
the possibility of exciting many particle-hole pairs
should be retained, in order
to describe the power-law singularities. In fact, this feature is
present in the parquet diagram approach
used by Abrikosov and Nozi\`eres and
coworkers \cite{ABRIKOSOV,NOZIERES1}.
The similarity between a parquet calculation and our simple
three-body approach arises because in the former
the three-particle vertex function involves
only successive two-particle
interactions. However, since the same parquet diagram
corresponds to several
time orderings for the interaction events,
the actual number of particle-hole
pairs is not fixed to unity. Unfortunately, our
simplified approach does not seem to provide a way to
bypass the cumbersome parquet algebra.
The main complication encountered upon going from the Faddeev approach, with
fixed particle number, to the parquet algebra in the full Hilbert space, is
reflected by the need to keep frequencies as
additional integration variables
in the internal lines of the graphs. For the X-ray
edge problem with a separable
potential, momentum integrations are staightforward and
we recover a one-dimensional
problem (in the time direction). For more realistic systems, such as the
two-dimensional Hubbard model, such a simplification is not so obvious.
This is why our simpler method may still be useful there, since it may
for instance be able to indicate the presence of spin and charge decoupling.
We hope to be able to address this issue in a forthcoming investigation.

\acknowledgements
The authors wish to thank J. Voit for discussions concerning
exact Green functions of the Luttinger model.

\figure{
Diagrams in the one-electron self-energy
accounted for in the Faddeev Equations.
As opposed to the usual convention the vertices
are time ordered from left to right. This prevents a violation of
the variational assumption of the presence of at most three excitations
at one time. The middle line is the spin up electron and the outer lines
are the spin down particle-hole pair.
\label{DIAGRAM}
}
\figure{
One branch Luttinger model: comparison of
Im~G for the exact solution (dashed line) and the
exact solution of the three-body approximation
(solid line)
as a function of normalized energy
$x = ({\tilde\omega}-v_{F}{\tilde q})/U{\tilde q}$
for fixed total momentum $q$.
\label{ONEBRANCH}
}
\figure{
Two branch Luttinger model:
Im~G as a funtion of normalized energy
${\tilde\omega}/v_{F}{\tilde q}$.
Solid line: three-body approximation;
Short dashed line: exact solution;
Long dashed line: three-body approximation reflected
about its maximum.
The non-zero part of Im~G at negative energies is not shown.
\label{TWOBRANCH}
}
\figure
{
Two branch Luttinger model: Real part of the self-energy
defined as the
energy shift (relative to and normalized by the
non-interacting quasi-particle
energy $v_{F}{\tilde q}$) of the peak of Im~G
as a function of the interaction strength $U/v_{F}$.
Dashed line: exact solution; Diamonds: three-body approximation.
\label{SHIFT}
}
\figure
{
Spectral density of the on-site particle (solid line)
and hole (dashed line) Green functions in the presence
of the deep hole. The values $k_{F}/D = 0.1$ and
$V/Dv_{F} = 0.2$ have been used.
\label{IMG}
}
\figure
{
Self energy of the deep level as a function of
frequency using the same parameters as
in Fig. \ref{IMG}. Real part: solid line,
Imaginary part: dashed line.
\label{SELF}
}
\end{document}